\documentclass[twocolumn]{aastex63}

\usepackage{soul}

\hypersetup{linkcolor=red,citecolor=green,filecolor=cyan,urlcolor=magenta}

\def\src {RRAT~J1918$-$0449~}
\newcommand{\pcm}{\,pc\,cm$^{-3}$}	

\received{XXX}
\revised{XXX}
\accepted{XXX}
\submitjournal{ApJ}

\shorttitle{\src discovered at FAST}
\shortauthors{J. L. Chen et al.}

\begin{document}

\title{The discovery of a rotating radio transient J1918$-$0449 with
intriguing emission properties with the five hundred meter aperture spherical
radio telescope}

\correspondingauthor{Z. G. Wen}
\email{wenzhigang@xao.ac.cn}

\author{J. L. Chen}
\affiliation{Department of Physics and Electronic Engineering, \\
Yuncheng University, Yuncheng, Shanxi, 044000, People's Republic of China}
\affiliation{Xinjiang Astronomical Observatory, Chinese Academy of Sciences, \\
150, Science-1 Street, Urumqi, Xinjiang, 830011, People's Republic of China}

\author{Z. G. Wen}
\affiliation{Xinjiang Astronomical Observatory, Chinese Academy of Sciences, \\
150, Science-1 Street, Urumqi, Xinjiang, 830011, People's Republic of China}
\affiliation{Key laboratory of Radio Astronomy, Chinese Academy of Sciences, \\
Nanjing, 210008, People’s Republic of China}
\affiliation{Key Laboratory of Microwave Technology, \\
Urumqi, Xinjiang, 830011, People’s Republic of China}
\affiliation{Guizhou Provincial Key Laboratory of Radio Astronomy and Data Processing, \\ 
Guiyang, Guizhou, 550001,People’s Republic of China}

\author{J. P. Yuan}
\affiliation{Xinjiang Astronomical Observatory, Chinese Academy of Sciences, \\
150, Science-1 Street, Urumqi, Xinjiang, 830011, People's Republic of China}
\affiliation{Key laboratory of Radio Astronomy, Chinese Academy of Sciences, \\
Nanjing, 210008, People’s Republic of China}

\author{N. Wang}
\affiliation{Xinjiang Astronomical Observatory, Chinese Academy of Sciences, \\
150, Science-1 Street, Urumqi, Xinjiang, 830011, People's Republic of China}
\affiliation{Key laboratory of Radio Astronomy, Chinese Academy of Sciences, \\
Nanjing, 210008, People’s Republic of China}

\author{D. Li}
\affiliation{National Astronomical Observatories, Chinese Academy of Sciences, \\ 
Beijing, 100101, People’s Republic of China}
\affiliation{Xinjiang Astronomical Observatory, Chinese Academy of Sciences, \\
150, Science-1 Street, Urumqi, Xinjiang, 830011, People's Republic of China}
\affiliation{NAOC-UKZN Computational Astrophysics Centre, University of KwaZulu-Natal, \\ 
Durban, 4000, South Africa}

\author{H. G. Wang}
\affiliation{School of Physics and Electronic Engineering, \\
Guangzhou University, 510006, Guangzhou, People's Republic of China }
\affiliation{Xinjiang Astronomical Observatory, Chinese Academy of Sciences, \\
150, Science-1 Street, Urumqi, Xinjiang, 830011, People's Republic of China}

\author{W. M. Yan}
\affiliation{Xinjiang Astronomical Observatory, Chinese Academy of Sciences, \\
150, Science-1 Street, Urumqi, Xinjiang, 830011, People's Republic of China}
\affiliation{Key laboratory of Radio Astronomy, Chinese Academy of Sciences, \\
Nanjing, 210008, People’s Republic of China}

\author{R. Yuen}
\affiliation{Xinjiang Astronomical Observatory, Chinese Academy of Sciences, \\
150, Science-1 Street, Urumqi, Xinjiang, 830011, People's Republic of China}

\author{P. Wang}
\affiliation{National Astronomical Observatories, Chinese Academy of Sciences, \\ 
Beijing, 100101, People’s Republic of China}

\author{Z. Wang}
\affiliation{School of Physical Science and Technology, \\
Xinjiang University, urumqi, Xinjiang, 830046, People's Republic of China}
\affiliation{Xinjiang Astronomical Observatory, Chinese Academy of Sciences, \\
150, Science-1 Street, Urumqi, Xinjiang, 830011, People's Republic of China}

\author{W. W. Zhu}
\affiliation{National Astronomical Observatories, Chinese Academy of Sciences, \\ 
Beijing, 100101, People’s Republic of China}

\author{J. R. Niu}
\affiliation{National Astronomical Observatories, Chinese Academy of Sciences, \\ 
Beijing, 100101, People’s Republic of China}

\author{C. C. Miao}
\affiliation{National Astronomical Observatories, Chinese Academy of Sciences, \\ 
Beijing, 100101, People’s Republic of China}

\author{M. Y. Xue}
\affiliation{National Astronomical Observatories, Chinese Academy of Sciences, \\ 
Beijing, 100101, People’s Republic of China}

\author{B. P. Gong}
\affiliation{Department of Physics, Huazhong University of Science and Technology, \\
Wuhan, 430074, People’s Republic of China}

\begin{abstract}
In this study, we report on a detailed single pulse analysis of the radio
emission from a rotating radio transient (RRAT) J1918$-$0449 which is the first 
RRAT discovered with the five hundred meter aperture spherical radio telescope (FAST).
The sensitive observations were carried out on 30 April 2021 using the FAST
with a central frequency of 1250 MHz and a short time resolution of 49.152
$\mu$s, which forms a reliable basis to probe single pulse emission properties
in detail.
The source was successively observed for around 2 hours.
A total of 83 dispersed bursts with significance above 6$\sigma$ are detected 
over 1.8 hours.
The source's DM and rotational period are determined to be 116.1$\pm$0.4 \pcm \ 
and 2479.21$\pm$0.03 ms, respectively.
The share of registered pulses from the total number of observed period is
3.12\%.
No underlying emission is detected in the averaged off pulse profile.
For bursts with fluence larger than 10 Jy ms, the pulse energy follows a
power-law distribution with an index of $-3.1\pm0.4$, suggesting the existence
of bright pulse emission.
We find that the distribution of time between subsequent pulses is consistent with 
a stationary Poisson process and find no evidence of clustering over the 1.8 h
observations, giving a mean burst rate of one burst every 66 s.
Close inspection of the detected bright pulses reveals that 21 pulses exhibit 
well-defined quasi-periodicities.
The subpulse drifting is present in non-successive rotations with periodicity of
$2.51\pm0.06$ periods.
Finally, possible physical mechanisms are discussed.
\end{abstract}

\keywords{pulsars: individual (\src)}

\section{Introduction}
\label{sec:int}
Rotating radio transients (RRATs) are characterized by short radio bursts with a
typical duration of a few milliseconds.
The bursts occur sporadically at irregular and infrequent intervals of minutes
to hours, which results in undiscovered periodicities using standard Fourier or 
periodogram techniques \citep{Keane+etal+2011}.
The nulling fraction of RRATs can exceed 99 per cent. 
The dedicated timing observations of the bursts reveal underlying
regularity of the order of seconds and they have comparable spin-down rates to
canonical pulsars \citep{Mclaughlin+etal+2006}.
The detection of the thermal X-ray spectrum from RRAT J1819$-$1458 further 
illustrates the true nature of the star as a cooling neutron star 
\citep{Reynolds+etal+2006}.
The average magnetic fields of RRATs and the averaged period derivatives are 
higher than those of ordinary pulsars \citep{Cui+etal+2017}.
The log-normal pulse energy distribution and power-law frequency dependence of
the mean flux are both consistent with those for the rotation powered 
pulsars \citep{Keane+McLaughlin+2011}.
However, the nature of RRAT's emission is still to be fully understood.
Potential links between RRATs and the other pulsar populations have been
identified.
A post-glitch over-recovery in the frequency derivative for RRAT J1819$-$1458
was found by \citet{Bhattacharyya+etal+2018}, which typifies a magnetar.
PSR B0656+14 is a nearby middle-aged pulsar with strong pused high-energy
emission \citep{Weltevrede+etal+2010}.
In the radio band, it is possibly classifiable as a RRAT, because the profile
instabilities are observed to be caused by very bright and narrow pulses
\citep{Weltevrede+etal+2006b}.
Additionally, RRATs exhibit an extreme nulling phenomenon, staying in the quiet
state for most of the time.
The high nulling fractions in pulsars are speculated to be more related to
large characteristic age than long period \citep{Wang+etal+2007}.
In order to further probe the connection between RRATs and other neutron star
populations and the nature of the emission, the discovery of additional RRATs is
required.
Since the serendipitous discovery of RRAT in a search for isolated radio bursts 
in the Parkes multibeam pulsar survey data \citep{Mclaughlin+etal+2006}, a total
of 112 RRATs\footnote{\url{http://astro.phys.wvu.edu/rratalog}} have been 
detected at constant DM so far with identified periodicities in the range 
0.125 $-$ 7.7 s.

The RRATs were initially thought to be a distinct population of Galactic neutron stars. 
However, a problem is raised that the Galactic supernova rate appears to be 
insufficient to account for the inferred number of neutron stars
\citep{Keane+Kramer+2008}.
An evolutionary link between the whole population of neutron stars was proposed to be 
satisfactory to resolve this problem \citep{Keane+etal+2011}.
And the RRATs probably settle into a stable existence on the neutron star evolution.
Up to present, the neutron star spin evolutionary framework has not been established.
The discovery of additional RRATs help bridge populations in the so-called neutron star
zoo in an attempt to understand their origins and evolution.
To further reveal the nature of RRAT emission and uncover connections between RRATs and
canonical pulsar population, it is also important to investigate their pulse energy
distributions, timing periodicities and pulse clustering, microstructure, and 
subpulse modulation.
The radio pulses from RRATs are detected at rates as low as once per three hours
to as frequently as one every few minutes \citep{Mclaughlin+etal+2006}.
The question of interest is whether the sources are truly `off' during the
rotation periods where the emission ceases suddenly or whether they still have some weak
underlying emission.
The FAST sensitivity allow for unprecedented potential to discover a substantial
proportion of RRATs, and thus shed insight into their physical emission mechanism and the
evolutionary paths of neutron stars.

\src is the first RRAT discovered in the commensal radio astronomy FAST survey
\citep[CRAFTS\footnote{\url{http://crafts.bao.ac.cn}};][]{Nan+etal+2011,Li+etal+2018,Qian+etal+2019,Zhang+etal+2019,Cameron+etal+2020,Cruces+etal+2021,Wangs+etal+2021,Wangsq+etal+2021}
using a single pulse search technique \citep{Zhu+etal+2014}.
In this paper, we present the discovery and analysis of the intriguing emission behavior
of \src observed with the FAST, which is organized as follows.
The observations are discussed in Section~\ref{sec:obs} along with description
of the data reduction methods. 
In Section~\ref{sec:res}, we present the results, namely the determination of the 
rotational period and dispersion measure from detected bursts, and the analysis of 
the emission properties.
The implication of our findings are discussed in Section~\ref{sec:dis}.
Section~\ref{sec:con} summarizes the results and discussion.

\section{Observations and data processing}
\label{sec:obs}
\src was originally discovered with the FAST on 2018 August 9 in the drift-scan 
survey using a 19-beam receiver to search for bright single pulses.
The dispersion measure (DM) was measured to be 119 \pcm.
The follow-up observations of \src were carried out with the FAST on 2021 April 30 at 
20:37:59 UT using the 19-beam receiver with frequency covering from 1000 to 
1500 MHz \citep{Jiang+etal+2020}.
The orthogonal linear polarizations from all beams were received and amplified 
using a low-noise cryogenic amplifier in the cabin and then transferred to the 
control room through fibers.
At the control room, the radio frequency signals were down
converted using a local oscillator with frequency of 1 GHz, producing intermediate
frequency (IF) signals with bandwidth of 500 MHz.
The properly-amplified IF signals were then sent to the pulsar machine, which was 
developed on a ROACH2\footnote{\url{https://casper.berkeley.edu/wiki/ROACH2}}
platform, for further processing.
The IF signals were converted into raw voltage signals and digitized using
the high speed analog-to-digital converters.
The digital signals were then processed on the field programmable gate array (FPGA).
Across the entire 500 MHz passband, 1024 frequency channels were produced, using
a polyphase filterbank, and correlated to compute the coherency matrices.
The channelized data were integrated in time for 49.152 $\mu$s per spectrum
before recording in a search mode PSRFITS format \citep{Hotan+etal+2004} with 8 
bit quantization.
The source was successively observed for around 2 hours.

In the off-line data processing, the raw filterbank data were initially processed 
to mitigate spurious radio frequency interference (RFI).
Following the technique described by \citet{Wen+etal+2021}, the 
bandpass was removed by dividing each sample in a channel by the average value in
that channel as calculated over 2 s intervals.
The narrowband and broadband RFI was identified and mitigated by deploying the
statistical method suggested by \citet{ter+etal+2019}.
Subsequently, incoherent dedispersion was applied to correct for the dispersive 
delay induced by free electrons on the arrival time of radio waves.
To determine the accurate value of DM, a search over this parameter ranging from
100 to 130 \pcm \ in steps of 0.1 \pcm \ was performed.
Each of these dispersive corrected time series was searched for significant
signals above a threshold, over different integration times, using a boxcar
profile.
The detection threshold was set to be 6$\sigma$, where $\sigma$ is the standard
deviation.

Since our observations were not absolutely flux calibrated based on a noise 
generator, to convert our data to a Jansky scale, the flux density ($S$) of 
each burst pulse was estimated using the radiometer equation 
\begin{equation}
	\centering
	S = \frac{(\rm{S/N}) S_{\rm{sys}}}{\sqrt{n_{\rm{p}} t_{\rm{int}} \Delta\nu}},
\end{equation}
where $S_{\rm{sys}}=2$ Jy is the system equivalent flux density, $n_{\rm{p}}=2$ is the
number of polarizations, $t_{\rm{int}}=49.152\ \mu$s is the integration time, and
$\Delta\nu=500$ MHz is the observing bandwidth.
It is worth noting that this equation is only applicable for pulses that are
longer than the time resolution.
Our detection threshold corresponds to a flux density of 54.13 mJy for a minimum
signal-to-noise ratio (S/N) of 6.

\section{Results}
\label{sec:res}

\subsection{Burst detection}
A total of 83 burst pulses were detected in only a single beam (Beam 15) of the
19-beam receiver using the HEIMDALL single pulse processing
software\footnote{\url{http://sourceforge.net/projects/heimdall-astro}}.
The remaining 18 beams show no clear candidates, nor RFI around the times of the
bursts.
Therefore, the precise position of the source cannot be provided by fitting the
relative signal strength in each beam to the side-lode pattern of the multiple
beams \citep{Ravi+etal+2016}.
The pointing position of the Beam 15 at the reference epoch with J2000 coordinates is
simply derived to be: right ascension = $\rm 19^h18^m36^s.07$ and declination =
$-04^\circ49'49''.80$, according to the angular distance of the beam center to
that of the central beam M01 \citep{Jiang+etal+2020}.
The topocentric arrival times of detected bursts were recorded, along with other
properties including S/N, DM, and matched filter width.
In order to reject the false positives, a visual inspection was carried out on
the generated diagnostic plots.

\begin{figure}
	\centering
	\includegraphics[width=8.0cm,height=8.0cm,angle=0,scale=1.0]{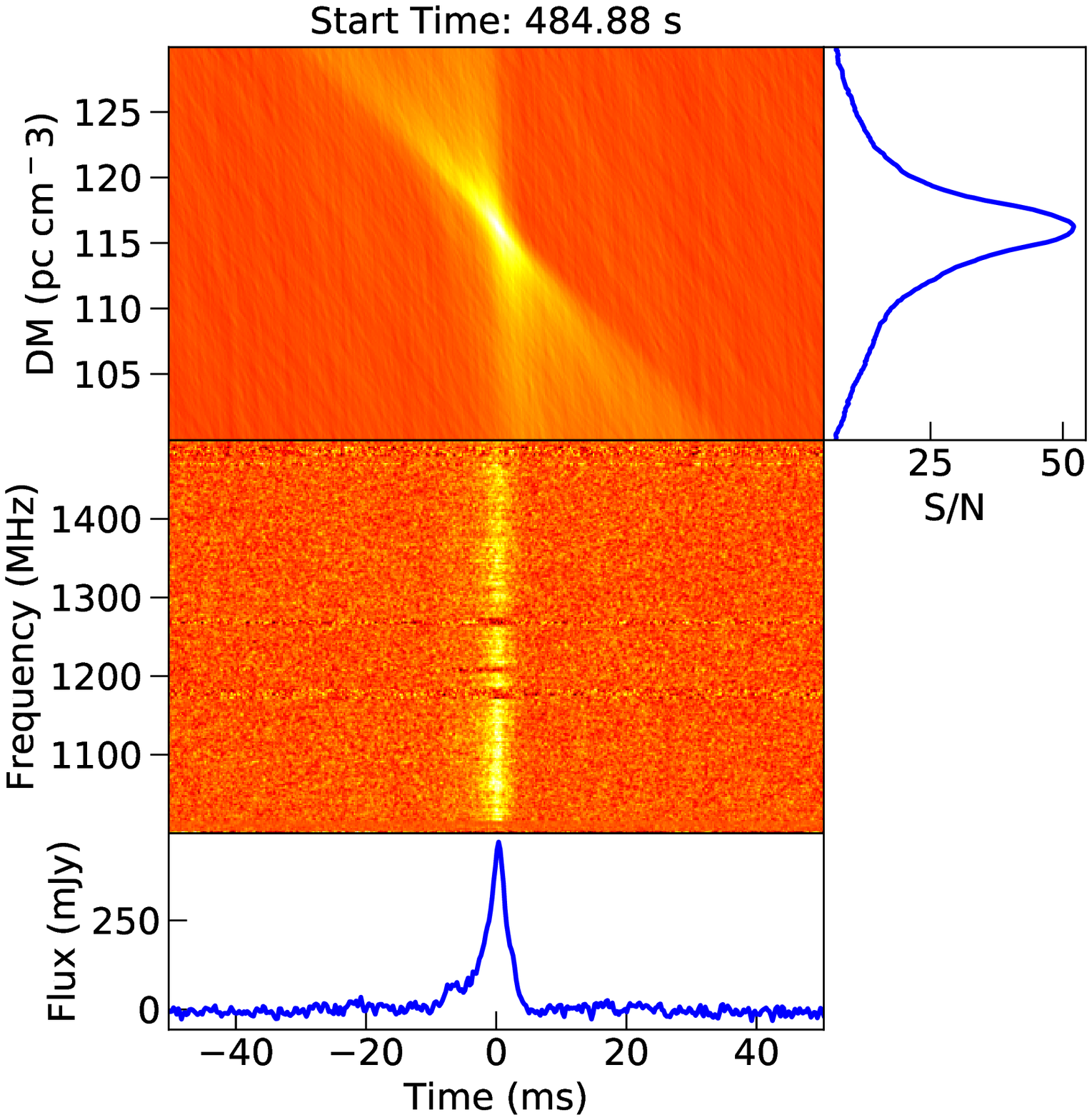}
	\caption{Diagnostic plot showing a detected burst pulse from \src.
	The upper-left panel is a DM-time color-coded diagram generated by de-dispersing 
	the signal with DM varying from 100 to 130 \pcm \ in steps of 0.1 \pcm.
	The upper-right panel is a DM-S/N graph, in which the S/N for each DM is calculated
	from the bin of maximum amplitude over 100 ms.
	The middle panel presents the waterfall plot of the dedispersed dynamic spectra 
	at DM=116.1 \pcm.
	No smearing is shown in the spectra, indicating that the value of DM is
	correct.
	The lower panel is a dedispersed time series graph showing the flux density of
	each bin, which clearly delineates a burst with single component profile.
	The time since the start of the observations is shown in the title.}
	\label{pic:burst_diagnostic}
\end{figure}

Figure~\ref{pic:burst_diagnostic} shows a diagnostic plot, containing a DM
versus time color-coded graph (upper-left panel), a S/N versus DM graph 
(upper-right panel), a waterfall diagram of dedispersed dynamic spectra 
(middle panel), and a dedispersed time series diagram (lower panel).
In DM-time space, a burst is detected as a vertical strip of pulses at multiple DMs
ranging from 110 to 120 \pcm.
The strip is gradually broadening and finally smears out with the DM either 
increasing or decreasing.
The burst is clearly visible in the DM versus time and the S/N versus DM
diagrams with maximum S/N around DM of 116.1 \pcm, indicating that it is
astrophysical.
There is no smearing shown in the waterfall image of dedispersed dynamic
spectra, which confirms that the value of DM is correct.
It is noted that there is no single pulse exhibiting the same narrow-band frequency
structure that is shown in FRB 121102 \citep{Gajjar+etal+2018}.
The lower panel shows its integrated pulse profile after being dedispersed at
the optimal DM value.
The corresponding flux density scale is applied to the single-pulse time series.

In order to determine the DM of this source, the averaged S/N versus DM diagram
is produced by averaging all S/Ns in each DM step of all 83 bursts.
The mean value of DM is estimated to be $116.1\pm0.4$ \pcm \ , which is adopted
for dedispersion in the following analysis.
The DM-based distance is estimated to be 3.72 kpc from the NE2001 model for the 
Galactic free electron distribution \citep{Cordes+Lazio+2002}.


\subsection{Pulse energy and width}
\begin{figure}
	\centering
	\includegraphics[width=8.0cm,height=6.0cm,angle=0,scale=1.0]{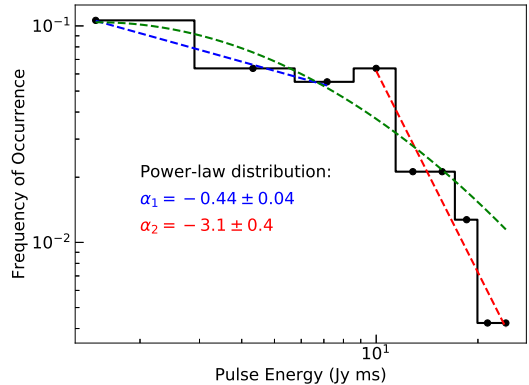}
	\caption{Distribution of energies at logarithmic scale for 83 bursts detected 
	from \src at 1250 MHz, which is best fitted with a broken power-law.
	The break of the power-law occurs at $\sim$10 Jy ms.
	The power-law indices are estimated to be $\alpha_1=-0.44\pm0.04$ (blue) and
	$\alpha_2=-3.1\pm0.4$ (red) for low and high energies, respectively.
	The best-fit log-normal distribution is indicated with green dashed curve as
	well.}
	\label{pic:fluence_dist}
\end{figure}

The pulse energy distribution provides a meaningful way to characterize the
frequency of occurrence of bursts, as well as a window into the state of the
plasma in the pulsar magnetosphere.
In order to study the pulse energy distribution, the fluence is measured by 
integrating all the flux densities within a time interval of each detected burst.
The most energetic pulse has a width of 21.09 ms, a fluence of 25.61 Jy ms,
and a peak flux density of 453.55 mJy.
The fluence distribution of all burst pulses is shown Figure~\ref{pic:fluence_dist}.
It is evident that a broken power-law fitting is necessary with $R$-square of
0.99.
The break of the power-law occurs at $\sim$10 Jy ms.
The power-law indices of the fluence distribution are determined to be
$\alpha_1=-0.44\pm0.04$ and $\alpha_2=-3.1\pm0.4$ for low and high energies,
respectively.
In addition we fit the pulse energy distribution with a pure log-normal
model, giving $R$-square of 0.89.

\begin{figure}
	\centering
	\includegraphics[width=9.0cm,height=8.0cm,angle=0,scale=1.0]{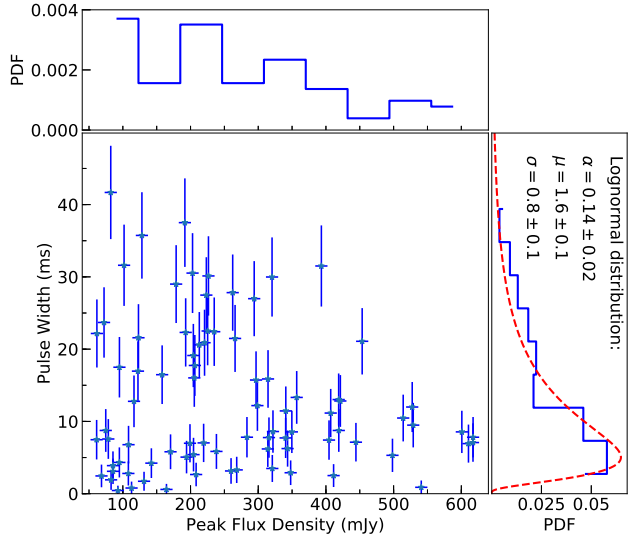}
	\caption{Main panel: scatter plot of pulse width and peak flux density for the 83
	detected bursts.
	Top panel: PDF of peak flux densities.
	Right panel: PDF of pulse widths along with the best-fitted logarithmic
	normal distribution.
	The best-fitting parameters are shown in the legend.}
	\label{pic:flux_width}
\end{figure}

The main panel of Figure~\ref{pic:flux_width} shows a scatter plot of peak flux
density and the pulse width for all detected bursts.
The uncertainty of the peak flux density is estimated using the rms value within the
off-pulse window.
The probability density function (PDF) for the peak flux densities is shown in the 
upper panel.
The pulse widths are determined by convolving each individual burst with a
series of boxcar filters with varying widths.
The matched boxcar filter size is calculated as the pulse width when the best
fit produces the optimum response.
The error of the pulse width is simply determined by the square root of its value.
The pulse width PDF is given in the right panel, which clearly presents a logarithmic 
normal distribution centered at $\sim$5.1 ms.

\subsection{Waiting-time distribution}
\begin{figure}
	\centering
	\includegraphics[width=8.0cm,height=6.0cm,angle=0,scale=1.0]{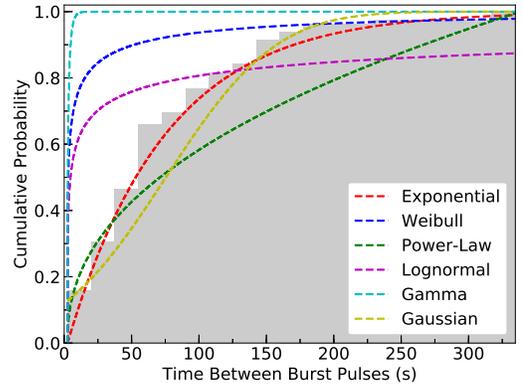}
	\caption{Normalized cumulative distribution of the time interval between
	successive burst pulses.
	The best-fit exponential distribution with $\lambda=0.015\pm0.003$ is shown 
	in the red dashed curve.
	In addition we fit the distribution with a pure Weibull (blue), power-law
	(green), Gaussian (yellow), log-normal (magenta) and Gamma (cyan), 
	respectively, which cannot describe the observed distribution well.
}
	\label{pic:wtd}
\end{figure}

The burst pulses from \src seem to happen in an irregular fashion.
The distribution of intervals between subsequent pulses (wait times) may contain 
hints to the emission mechanism.
Figure~\ref{pic:wtd} shows the cumulative density function of the these wait
times.
An exponential reduction in the separation time between the pulses is presented,
The arrival time separation may follow a Poisson process by assuming that the 
bursts are mutually exclusive events independent of each other.
The probability of a burst occurring in the interval $\Delta t$ is described as
$P(\Delta t) = \lambda e^{-\lambda \Delta t}$, where $\lambda$ is the mean
occurrence rate.
To quantify the fit, the Kolmogorov-Smirnov statistics (K-S test) of the sample
versus its exponential distribution fit is performed.
The obtained $p-$value is greater than the threshold value of 0.05, which indicates
that the observed distribution is consistent with a exponential distribution.
The waiting-time distribution is fitted to the exponential model using a 
least-square fitting method, which is shown in the red dashed curve.
The function with an exponential decay with time constant determined to be
$\lambda=0.015\pm0.003$ is in excellent agreement with the observed data.
The mean burst rate is estimated to be one burst every 66 s observed above our
threshold limit.
We also attempt to model this population as a pure Weibull, power-law, Gaussian,
log-normal and Gamma, respectively, which cannot describe the
observed distribution well.
In addition, there is no significant correlation between the wait time and the 
fluence of the detected pulses, which implies that the emission is not a
consequence of the process through which the energy is stored up, such as 
magnetic recombination in the pulsar magnetosphere \citep{Lyutikov+2002}.

\subsection{Rotational period determination}
\begin{figure}
	\centering
	\includegraphics[width=8.0cm,height=6.0cm,angle=0,scale=1.0]{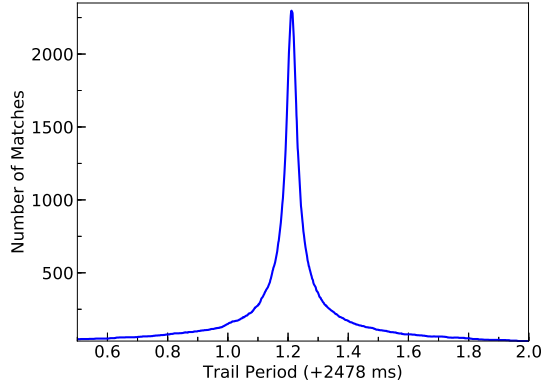}
	\caption{Periodogram of \src.
	A distinct peak of the cumulative number of pulsation presence is located at
	$P=2479.21\pm0.03$ ms.}
	\label{pic:period}
\end{figure}

The RRAT emission has an underlying periodicity, even though the interval
between individual pulses varies widely.
The spin-period of a particular RRAT is generally determined by finding the 
greatest common denominator of the intervals between pulses.
Due to the limited number of pulses, a multiple of the period is generated 
instead of the true period of the neutron star by using this method.
To search for the periodic pulsations from \src, the burst arrival
times are folded and grouped with different periods from 0.1 to 50 s in steps of
0.01 ms using the epoch folding technique described by \citet{Leahy+etal+1983}.
Figure~\ref{pic:period} shows the number of matches with tolerance of 10\% in
phase.
It is noticed that a distinct peak is shown at $P_1=2479.21\pm0.03$ ms,
corresponding to the rotational period of the neutron star.
The calculated error corresponds to the FWHM of the peak in the periodogram.

\subsection{Pulse shapes}

\begin{figure}
\centering
	\includegraphics[width=6.0cm,height=8.0cm,angle=0,scale=1.0]{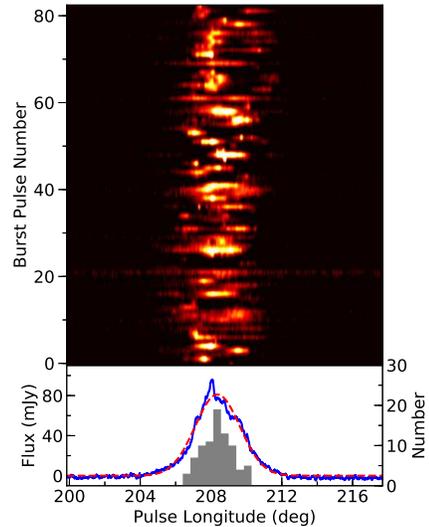}
	\caption{Top panel: detected burst pulses from RRAT J1918$-$0449 are folded
	at the 2479.21 ms period at a centre frequency of 1250 MHz.
	Bottom panel: longitude-resolved flux density by integrating all bursts
	along with the longitude distribution of the detected burst pulses.}
	\label{pic:sgl}
\end{figure}

\begin{figure}
\centering
	\includegraphics[width=8.0cm,height=6.0cm,angle=0,scale=1.0]{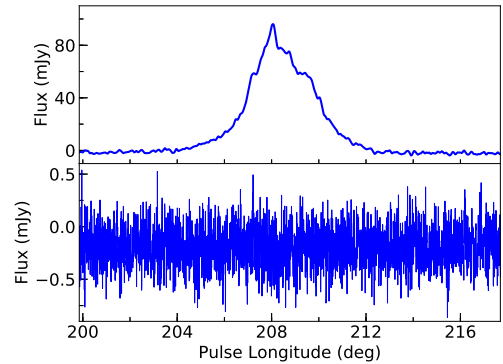}
	\caption{The null and burst pulse proﬁles obtained by integrating
	corresponding classified nulls and bursts.
	The null pulse proﬁle does not show any signiﬁcant emission.}
	\label{pic:prof}
\end{figure}

All detected burst pulses from the observations are extracted and folded at the
period of 2479.21 ms.
The upper panel of Figure~\ref{pic:sgl} displays the individual pulse sequence
as a colour-map plot, where the bursts are aligned in longitude.
The lower panel shows the histogram of the longitudes of the detected burst
pulses superimposed on the average pulse profile by integrating all burst 
pulses.
The pseudo-integrated profile shows the presence of a single component with 
measured FWMH of 17.45$\pm$0.03 ms (2.534$\pm$0.004) by fitting with a Gaussian
profile.
It is noted that the folded profile is not well defined by virtue of a small
sample of sporadic bursts.
The bursts all fall within the pulsar emission window, with a higher occurrence
preferred in the center of the pulse profile.
It is essential to confirm whether there are weak pulses with S/Ns lower than
the threshold that are not detectable by the single-pulse search pipeline.
Figure~\ref{pic:prof} presents the integrated pulse profiles obtained from
all classified bursts and nulls.
The weak pulse profile is constructed by subtracting all detected pulses from
the composite profile.
No emission in the weak pulse profile is detected with significance above
3$\sigma$.

\subsection{Microstructure}
The average pulse profile in Figure~\ref{pic:sgl} shows no obvious signatures of 
microstructure at any given longitude, since the averaging process washes out 
any such structure that can be seen in individual pulses. 
The sensitive observations with sufficiently high time resolution are performed 
with FAST, forming a reliable basis for detection of micropulses.
As shown in the upper panel of Figure~\ref{pic:sgl}, the presence of some pulses 
with narrow bursty features is evident.
The structure in single pulses varies dramatically on a timescale down to almost
the resolution interval from pulse to pulse.

\begin{figure*}
	\centering
	\includegraphics[width=16.0cm,height=8.0cm,angle=0,scale=1.0]{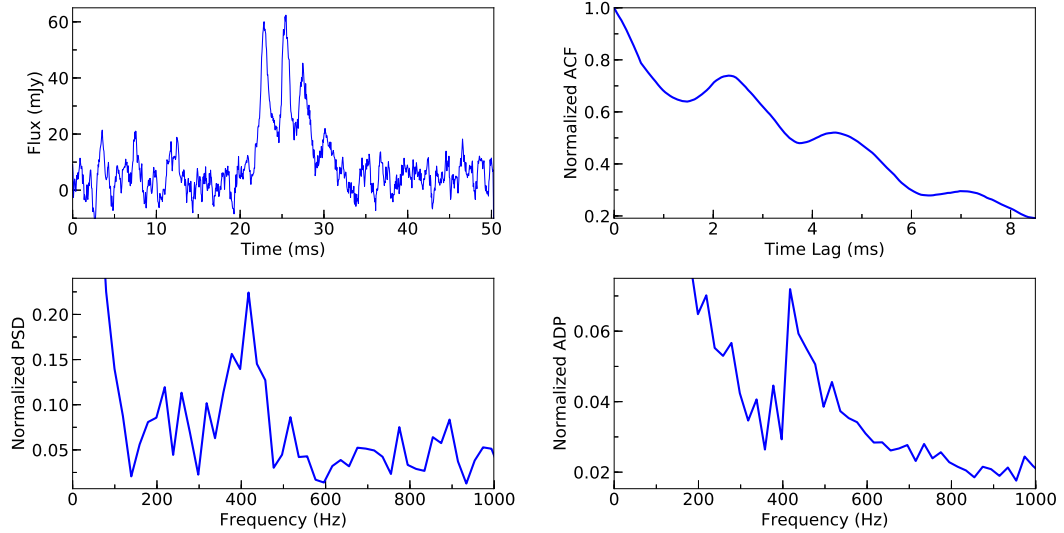}
	\caption{Example of a burst at 1250 MHz showing microstructure (top left) and 
its computed ACF (top right), PSD (bottom left), and ADP (bottom right).
Note the prominent oscillations in the single-pulse ACF produced by the quasi-periodic
microstructure. 
The quasi-periodicity in the micropulses is visible in the PSD and ADP.}
	\label{pic:micro_burst23}
\end{figure*}

To quantify the characteristic timescales of subpulses and micropulses, the
standard analysis procedure is the computation of autocorrelation functions (ACFs)
for detected bright pulses.
Furthermore, the power spectrum of the data (PSD) and the power spectrum of the ACF 
derivative (ADP) are calculated to determine quasi-periodicities \citep{Lange+etal+1998}.
Figure~\ref{pic:micro_burst23} shows a typical example of the microstructure analysis
on a detected burst from \src.
The ACF shows a variation in slope at the time lag corresponding to the micropulse
width of $\tau_\mu=1.47\pm0.25$ ms.
The typical subpulse width of $\tau_s=3.74\pm0.25$ ms is indicated with a further slope 
change, that is the second-order turn-off point.
The characteristic separation of subpulses of $P_2=4.47\pm0.25$ ms, particularly evident
for drifting subpulses, is given by the time lag of the peak of a second subpulse feature.
The uncertainty of timescales is calculated from the effective time resolution
$t_{\rm{eff}}=\sqrt{t^2_{\rm{samp}}+t^2_{\rm{DM}}}$, where $t_{\rm{DM}}=240.87\
\mu$s is the dispersion smearing in an interchannel.
The feature near zero time lag represents the autocorrelation of the noise
relative to the pulsar signal, which decorrelates on an inverse bandwidth time
scale.
The existence of a quasi-periodicity within the pulse is noticeable as equally
spaced peaks in the ACF, the time lag of the first peak indicates the
periodicity of $P_\mu = 2.31\pm0.25$ ms.
Furthermore, the frequency peaks in the PSD and ADP are recorded along with the
value found from the ACF analysis.

\begin{figure}
	\centering
	\includegraphics[width=8.0cm,height=10.0cm,angle=0,scale=1.0]{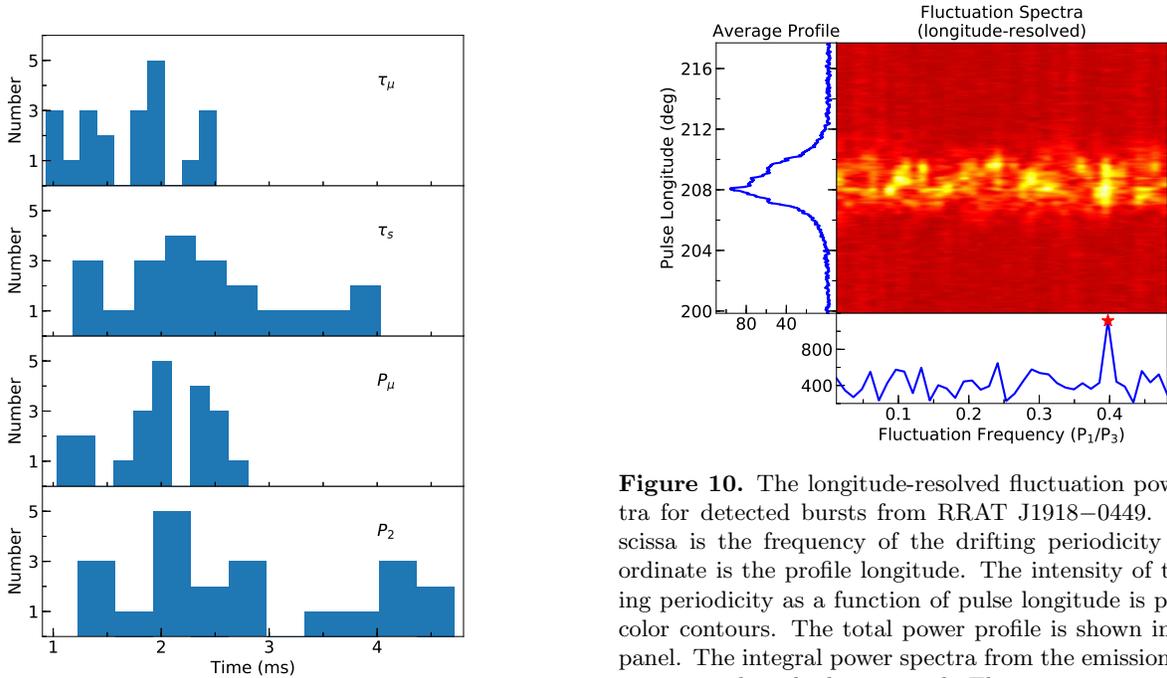}
	\caption{Distributions of micropulse width, subpulse width, microstructure
	quasi-periodicity, and subpulse separation for RRAT J1918$-$0449.}
	\label{pic:micro_hist}
\end{figure}

A careful visual inspection of the detected bursts and their corresponding ACF, 
PSD, and ADP were carried out to decide which of the pulses have periodicities 
that are strong enough to be evaluated.
The quasiperiodic microstructure features are detected in 21 individual pulses.
Nevertheless, no preferred width for the micropulses could be obtained from either
the averaged ACF or the width histogram.
Figure~\ref{pic:micro_hist} presents the calculated histograms of
micropulse width, subpulse width, microstructure quasi-periodicity, and subpulse
separation for RRAT J1918$-$0449.

\subsection{Subpulse drifting}
\begin{figure}
	\centering
	\includegraphics[width=8.0cm,height=6.0cm,angle=0,scale=1.0]{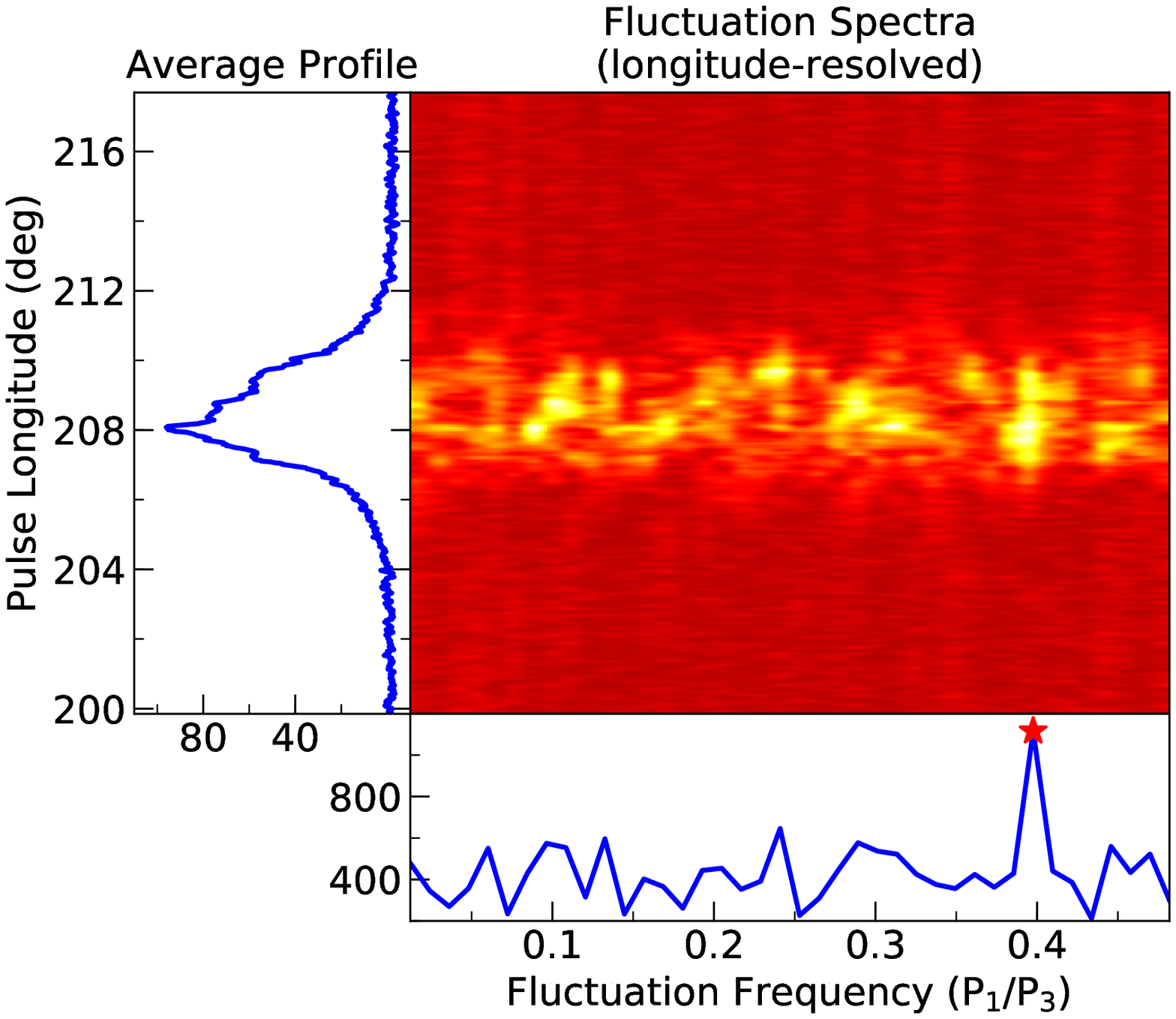}
	\caption{The longitude-resolved fluctuation power spectra for detected bursts
	from RRAT J1918$-$0449.
	The abscissa is the frequency of the drifting periodicity and the ordinate is
	the profile longitude.
	The intensity of the drifting periodicity as a function of pulse longitude 
	is plotted in color contours.
	The total power profile is shown in the left panel.
	The integral power spectra from the emission window is presented in the
	lower panel.
	The prominent modulation frequency is indicated with a red star.}
	\label{pic:lrfs}
\end{figure}

It is well known that at least one-third of the pulsars show systematic drifting of 
subpulses across the emission window for normal pulse 
emission \citep{Weltevrede+etal+2006a,Weltevrede+etal+2007}.
To investigate whether the detected burst pulses from RRAT J1918$-$0449 present
similar subpulse modulation in a systematic or a disordered fashion, a pulse 
sequence is artificially formed from the discrete bursts in chronological order.
Following the technique described by \citet{Backer+1970}, the 
longitude-resolved fluctuation spectra (LRFS) is adopted to characterize the 
subpulse modulation.
The LRFS is computed by performing discrete Fourier transforms over each
individual longitude bins.
Figure~\ref{pic:lrfs} shows the spectral power of fluctuations as a function of
rotational longitude.
Peak frequency in the fluctuation spectra represents periodicity of subpulse repetition at
any given pulse longitude, which is expressed as $P_1/P_3$.
It is noted that a distinct region of the LRFS with excess of power is confined 
in the emission window, which indicates that the bursts from the RRAT is modulated 
with a frequency of 0.40$\pm$0.01 cycles/period or a cycle of 2.51$\pm$0.06
rotational periods.
The periodicity and its corresponding uncertainty are estimated by fitting a Gaussian
function with 90\% confidence limit on the observed spectral line.

\begin{figure}
	\centering
	\includegraphics[width=8.0cm,height=6.0cm,angle=0,scale=1.0]{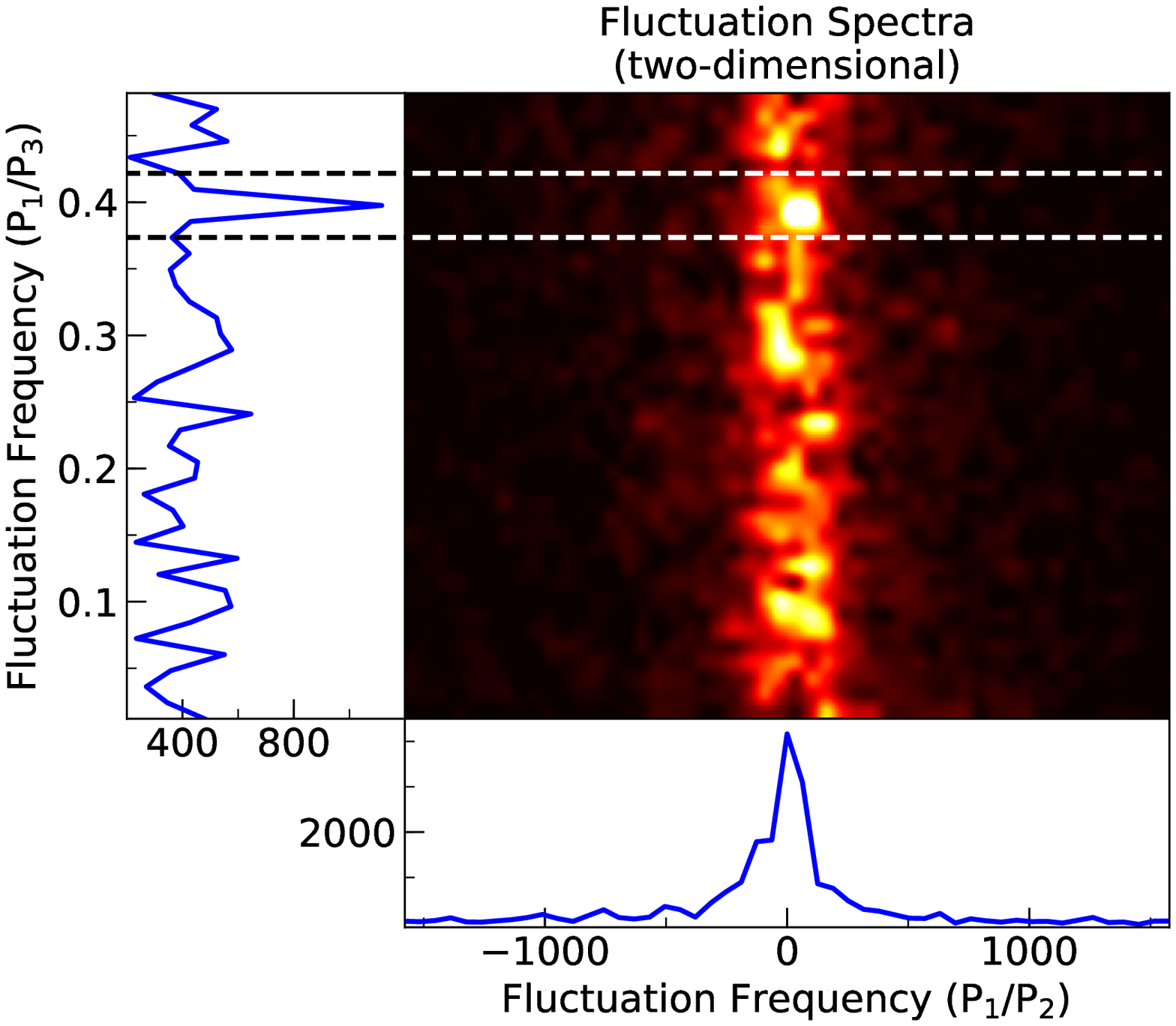}
	\caption{The two-dimensional fluctuation power spectra for detected bursts
	from RRAT J1918$-$0449.
	The units of the horizontal axis are cycles per period, which corresponds to
	$P_1/P_2$ (where $P_2$ is the horizontal drift band separation in time units
	in the case of drifting in longitude.)
	The power in the 2DFS is (between the dashed lines) horizontally and vertically
	integrated, producing the left and bottom panels.}
	\label{pic:twodfs}
\end{figure}

Although, the LRFS is a sensitive method to determine the time-averaged properties 
of periodic subpulse modulation, it is still in a state of uncertainty whether
the above determined modulation is originated from intensity or phase variation. 
In order to turn uncertainty into reality, the two-dimensional fluctuation spectrum
(2DFS) is calculated \citep{Edwards+Stappers+2002}. 
Following the same procedure used while calculating the LRFS, the
two-dimensional discrete Fourier transform is carried out on the emission
window (shown in Figure~\ref{pic:twodfs}).
The pattern repetition frequency along the pulse longitude is denoted in the 
horizontal axis of the 2DFS, which is expressed as $P_1/P_2$.
The value of $P_2$ is measured to be infinite since no significant phase offsets
corresponding to the peak frequency are presented, which may suggest that the 
emission is longitude stationary across the pulse window but periodically changes in
intensity.
Furthermore, the drifting is probably confined within a narrow phase region,
which is insensitive to the 2DFS technique.
As noted from the analysis of quasi-periodic microstructure, the characteristic
separation of subpulses is measured to be 0.65$\pm$0.04 degrees, which is much
narrower than the pulse profile.

In order to rule out the possibility that the occasional occurrence of strong subpulses
are dominating the spectra and therefore lead to misleading conclusions, a similar
method is carried out to further demonstrate the authenticity of periodic
fluctuations \citep{Weltevrede+etal+2006a}.
The order of the pulse sequence is randomized and then the LRFS is calculated
from the newly formed pulse stack.
No well-defined drifting periodicity in this process is detected in the 
emission window and noise window, which proves the significance of drift features.

\section{Discussion}
\label{sec:dis}

The sporadic nature of the emission from RRATs remains unclear currently, even 
several conjectures linked to intermittent pulsars and the nulling phenomenon 
have been put forward. 
\citet{Li+2006} suggested that RRATs can be extinct pulsars that are re-activated due 
to the interaction between the neutron star magnetosphere and the fallback of 
material from a supernova debris disk.
The asteroidal or circumpulsar debris that migrates into the pulsar light
cylinder before evaporating, ionizing, and perturbing particle acceleration
regions can disrupt current flows and electromagnetic radiation 
\citep{Cordes+Shannon+2008}.
Alternatively, RRATs are proposed to represent an extreme form of pulse nulling
within the framework of the standard canonical pulsar model \citep{Wang+etal+2007}.
A sharp change in the primary plasma density or the conditions of emission
propagation in the magnetosphere is proposed to be responsible for the character
of RRAT emission \citep{Timokhin+2010}.
It is shown that the profile changes are accompanied by changes in the spin-down
rate of the pulsar \citep{Kramer+etal+2006}, which further indicates the global 
changing in the magnetospheric configuration.

In the context of the general pulsar population, examples of clustering have
been noted for nulling \citep{Redman+Rankin+2009}, mode changing\citep{Wen+etal+2020} 
and bright pulses \citep{Wen+etal+2021}.
Additionally, the waiting-time distribution of FRB 121102 is well described by a
Weibull distribution \citep{Oppermann+etal+2018} as well, it is not as apt for 
the RRAT wait-time distributions.
The single-pulse emission is produced by a pure random process for RRATs 
J1913+1330 \citep{Shapiro-Albert+etal+2018} and J2325$-$0530 \citep{Meyers+etal+2019}, 
which are reasonably consistent with what we find for \src.
However, it is unclear whether this is the case for RRAT emission in general due
to the limited number of single-pulse detections.

Generally, the pulse energy distribution is considered as a useful metric for
comparing different pulsar emission modes.
The standard pulsar emission often results in log-normal 
distributions \citep{Burke-Spolaor+etal+2012}, while the so-called giant radio
pulses (GRPs) exhibit power-law distributions and are thought to be originated 
from a different emission mechanism \citep{Cordes+etal+2004}.
\citet{Cui+etal+2017} found that the single-pulse amplitude distributions of
RRATs could be described by either a log-normal, or a log-normal plus a 
power-law tail.
The constructed energy distribution of detected bursts for \src shows a broken
power-law distribution.
Considering the small sample of single pulses, we are unable to draw concrete
conclusions.
Ignoring the fact that the fluence cut-off is employed, the measured power-law
index is similar to that seen for Crab GRPs \citep{Bhat+etal+2008}.
However, the Crab GRPs are constrained in a narrow longitude range with
timescales on the order of a few microseconds and structure down to 10 
ns \citep{Hankins+etal+2003}, which do not follow the values found for \src.
We conclude that RRAT emission is not consistent with Crab GRP emission.

The rapid intensity fluctuations indicates a fundamental integral feature of 
pulsar radio emission \citep{Rickett+etal+1975}.
The investigation of micropulses and their quasi-periodicities are helpful to 
understand pulsar magnetospheric geometry and the longitudinally modulated 
emission pattern \citep{Benford+1977}.
The periodic effects may arise in the production and acceleration of charged
particles at the stellar surface \citep{Cheng+Ruderman+1980}.
\citet{Harding+Tademaru+1981} proposed that the quasi-periodicity in the 
microstructure is the result of radiative transfer effects in the 
magnetosphere.
An evident dependence between the pulsar rotational period and the micropulse 
width is identified, which indicates that the microstructure may arise from the 
sweeping of beams of emission \citep{Kramer+etal+2002}.

The phenomenon of subpulse drifting was first noted in the earliest days of
pulsar research \citep{Drake+Craft+1968}, which has historically been strongly
linked to magnetospheric phenomena, and thence, to the as-yet poorly understood
radio emission mechanism.
\citet{Ruderman+Sutherland+1975} proposed a carousel model in the seminal polar
gap theory.
The pattern of radio waves are characterized by a set of discrete pockets of
quasi-stable electrical activity (sparks) localized very near the pulsar surface.
The particles are accelerated through the polar gap region in the sites of sparks,
generating the secondary pair plasma. 
The curvature radiation are produced by an avalanche of secondary particles that
stream along the magnetic field lines at relativistic speeds. 
The electric potential along the magnetic field at the stellar surface are partially 
screened by the charged particles, which leads to the relative motion of the sparks
to the polar gap surface around the magnetic axis.
The magnetic azimuthal arrangement of sparks on the polar gap reflects the spatial
structure of the emission beam, which is constituted of discrete beamlets.
With each rotation, an Earth-bound observer perceives a different intensity pattern 
as the line of sight cuts through a slightly rotated emission cone.
However, the observed drifting in \src shows the systematic way that subpulses
arrive earlier in time than their predecessors (positive drifting in the classification 
scheme of \citet{Basu+etal+2019}) in non-successive rotations.
The information regarding the positions of the marching subpulses is somehow
stored during the nulls, and it is subsequently used to provide a starting point 
for the drifting pattern when emission of radio radiation recommences.
The persistence of drifting subpulse phase memory observed during pulsar nulling
has been noted in several pulsars.
\citet{Unwin+etal+1978} reported the particular behavior of subpulse phase during pulse
nulling observed from PSR B0809+74, in which the remarkable retention through a
null of the subpulse phase immediately prior to the onset of 
nulling.
After the null, the emission once again appears in drifting subpulses whose phase is
expected for the first null pulse.
Such memory behavior is also significant in PSR B0031$-$07 with longer nulling
duration than PSR B0809+74 \citet{Huguenin+etal+1970}.
For \src, the pulses are emitted randomly, which implies that the occurrence of 
bursts is restricted to specific portions of the integrated pulse window by the 
memory mechanism.

\section{Conclusions and summary}
\label{sec:con}
We have reported the first RRAT discovery of CRAFTS.
\src has a rotational period of 2479.21$\pm$0.03 ms, and a dispersion measure of
116.1$\pm$0.4 \pcm.
It is characterized by dispersed radio bursts of prominent duration 5.1 ms with one 
burst detected about every 1.3 min.
Even with the relatively small sample of single pulses presently at our
disposal, \src clearly exhibits many interesting single pulse behaviors, such
as exponential waiting-time distribution, quasi-periodic microstructure, and 
subpulse drifting in non-successive rotations.

The intriguing emission properties presented in \src indicate that RRATs are not
a homogeneous sample, but a mixture of different kinds of pulsars.
Further follow-up and multi-wavelength observations are required to carry out an
in-depth analysis of the timing, polarization properties of large sample of
RRATs, which are vital in understanding the physical emission mechanism.
The observations taken over many years has the added benefit of allowing us to
obtain statistically significant results on long timescales.
The characteristic age, surface magnetic field, spin-down luminosity and precise
position will be implied from the timing solutions.
The magnetospheric geometries can be constrained through high quality polarimetric 
profiles.

\section*{Acknowledgements}
\addcontentsline{toc}{section}{Acknowledgements}
We would like to thank the anonymous referee for providing constructive
suggestions to improve the article.
This work is partially supported by the National Natural
Science Foundation of China (NSFC grant No. U1631106, 11988101, U1838109, 12041304,
12041301, 11873080), the National SKA Program of China (2020SKA0120100),
and the Chinese Academy of Sciences Foundation of the young scholars of western 
(grant No. 2020-XBQNXZ-019).
Z.G.W. is supported by the 2021 project Xinjiang uygur autonomous region of China for
Tianshan elites. 
J.L.C. is supported by the Natural Science Foundation of Shanxi Province 
(20210302123083) and the Scientific and Technological Innovation Programs of Higher 
Education Institutions in Shanxi (grant No. 2021L470). 
D.L. is supported by the 2020 project of Xinjiang uygur autonomous region
of China for flexibly fetching in upscale talents.
W.M.Y. is supported by the CAS Jianzhihua project.
H.G.W. is supported by the 2018 project of Xinjiang uygur autonomous region of China 
for flexibly fetching in upscale talents. 
We thank members of the Pulsar Group at XAO for helpful discussions. 
This work made use of the data from the Five-hundred-meter Aperture Spherical radio
Telescope, which is a Chinese national mega-science facility, operated by National 
Astronomical Observatories, Chinese Academy of Sciences. 
We thank the observers in setting up the observations.

\bibliographystyle{aasjournal}
\bibliography{bibtex}

\end{document}